\documentclass[twocolumn,showpacs,preprintnumbers,amsmath,amssymb,superscriptaddress]{revtex4}
\usepackage[dvipdfmx]{color}
\usepackage[dvipdfmx]{graphicx}
\usepackage{dcolumn}
\usepackage{bm,amsmath,amssymb}

\begin{document}

\preprint{APS/123-QED}

\title{Study of photoinduced valence dynamics in EuNi$_2$(Si$_{0.21}$Ge$_{0.79}$)$_2$\\
through time-resolved X-ray absorption spectroscopy
}

\author{Y. Yokoyama$^*$}
\affiliation{Institute for Solid State Physics, University of Tokyo, Chiba 277-8581, Japan}
\affiliation{Department of Physics, University of Tokyo, Chiba 277-8561, Japan}
\author{K. Kawakami}
\affiliation{Graduate School of Engineering, Osaka Prefecture University, Sakai 599-8531, Japan}
\author{Y. Hirata}
\affiliation{Institute for Solid State Physics, University of Tokyo, Chiba 277-8581, Japan}
\affiliation{Department of Physics, University of Tokyo, Chiba 277-8561, Japan}
\author{K. Takubo}
\affiliation{Institute for Solid State Physics, University of Tokyo, Chiba 277-8581, Japan}
\author{K. Yamamoto}
\affiliation{Institute for Solid State Physics, University of Tokyo, Chiba 277-8581, Japan}
\affiliation{Department of Physics, University of Tokyo, Chiba 277-8561, Japan}
\author{K. Abe}
\affiliation{Graduate School of Engineering, Osaka Prefecture University, Sakai 599-8531, Japan}
\author{A. Mitsuda}
\affiliation{Graduate School of Science, Kyushu University, Fukuoka 819-0395, Japan}
\author{\linebreak H. Wada}
\affiliation{Graduate School of Science, Kyushu University, Fukuoka 819-0395, Japan}
\author{T. Uozumi }
\affiliation{Graduate School of Engineering, Osaka Prefecture University, Sakai 599-8531, Japan}
\author{S. Yamamoto}
\affiliation{Institute for Solid State Physics, University of Tokyo, Chiba 277-8581, Japan}
\affiliation{Department of Physics, University of Tokyo, Chiba 277-8561, Japan}
\author{I. Matsuda}
\affiliation{Institute for Solid State Physics, University of Tokyo, Chiba 277-8581, Japan}
\affiliation{Department of Physics, University of Tokyo, Chiba 277-8561, Japan}
\author{K. Mimura}
\affiliation{Graduate School of Engineering, Osaka Prefecture University, Sakai 599-8531, Japan}
\author{H. Wadati}
\affiliation{Institute for Solid State Physics, University of Tokyo, Chiba 277-8581, Japan}
\affiliation{Department of Physics, University of Tokyo, Chiba 277-8561, Japan}

\date{\today}

\begin{abstract}
The photoinduced valence dynamics of EuNi$_2$(Si$_{0.21}$Ge$_{0.79}$)$_2$ are investigated using time-resolved X-ray absorption spectroscopy for Eu $M_5$-edge.
Through the pump-probe technique with synchrotron X-ray and Ti:sapphire laser pulse,
a photoinduced valence transition is observed from Eu$^{3+}$ to Eu$^{2+}$.
Because the lifetime of a photoinduced state can be up to 3 ns, a metastable state is considered to be realized.
By comparing the experimental results with the theoretical calculations, the photoinduced valence transition between Eu 4$f$ and conduction electrons is quantitatively evaluated.
\end{abstract}

\maketitle

\newpage
The unique phenomena exhibited by 4$f$ electron systems, such as valence fluctuation, valence transition, and the Kondo effect, have drawn the attention of the scientific community. These phenomena originate from the hybridization between 4$f$ electrons and conduction electrons \cite{Varma}.
Recently, valence fluctuation and transition in Ce- and Yb-based compounds have reportedly exhibited unconventional superconductivity and non-Fermi-liquid behavior \cite{Jaccard,Trovarelli,Holmes1,Holmes2,Nakatsuji,Watanabe}.
In addition, the valence fluctuation has been related to quantum critical phenomena \cite{Nakatsuji2}.
Therefore, the mechanisms of valence fluctuation and transition are germane to the understanding of 4$f$ electron systems.

For the study of valence fluctuation and the valence transition, Eu compounds are one of the most suitable specimens due to their significantly large valence change.
Valence fluctuation and valence transition have been observed between Eu$^{2+}$ (4$f^7$, $J$ $=$ 7/2) and Eu$^{3+}$ (4$f^6$, $J$ $=$ 0) ions
\cite{Sampathkumaran,Croft,Segre,Wortmann,Mitsuda,Hesse,Wada1,Wada3,Wada2,Yamamoto1,Yamamoto2,Matsuda,Ichiki}.
The transition of Eu mean valence occurs by external stimuli such as temperature \cite{Sampathkumaran,Segre,Wortmann,Hesse,Wada1,Wada3,Wada2,Yamamoto1,Yamamoto2,Matsuda,Ichiki}, magnetic field \cite{Mitsuda,Wada1,Matsuda}, and/or pressure \cite{Hesse,Wada3}.
The detection of valence transitions was achieved using the Eu $M_5$-edge X-ray absorption spectroscopy (XAS)
\cite{Yamamoto1,Yamamoto2}.
However, the interaction of the compounds with photon irradiation is not fully understood.
Photon-controlled hybridization between the 4$f$ and conduction electrons may help realize a novel state and unravel more information on electron--photon interaction.
Therefore, in this study, the photoinduced dynamics of electronic structures were investigated.

Extensive studies on photoinduced transition in strongly correlated 3$d$ transition metal compounds were performed using pump-probe
X-ray spectroscopies, revealing dynamics such as insulator-to-metal transition
\cite{Radu,Beaud,Jal,Tsuyama,Takubo}.
These techniques involve controlling the time difference between a pump light, such as an visible laser pulse, and a probe light, such as a synchrotron radiation X-ray pulse,
to detect the time-dependent variation of photon-induced electronic structures.
In the soft X-ray region, the 3$d$ states were directly observed through the 2$p$ $\rightarrow$ 3$d$ resonance \cite{Tsuyama}.
As for the 4$f$ rare-earth compounds,
the spin states and the hybridization between 4$f$ electrons and conduction electrons
were studied via resonant X-ray diffraction with 3$d$ $\rightarrow$ 4$f$ resonance \cite{Nele} and reflectivity measurements \cite{Zhang}.
Although the valence of 4$f$ electron systems were clearly distinguished in X-ray absorption spectroscopy (XAS) spectra, the valence dynamics were not observed via XAS measurements.
Therefore, the photoinduced valence dynamics were investigated using time-resolved soft X-ray absorption spectroscopy (Tr-XAS).

For the study of photoinduced valence dynamics, EuNi$_2$(Si$_{1-x}$Ge$_{x}$)$_2$ with $x=0.79$ was selected as it exhibits a large valence change ($\sim$0.6), and the valence transition occurs at a relatively high temperature ($\sim$84 K) \cite{Ichiki}.
As the surface of EuNi$_2$(Si$_{0.21}$Ge$_{0.79}$)$_2$ is sensitive to oxygen, i.e., it tends to get oxidized,
the polycrystalline samples were fractured prior to the measurements in an ultra-high vacuum chamber at the possible lowest temperature ($\sim$40 K).
The static XAS and Tr-XAS measurements were performed at BL07LSU of SPring-8 \cite{Takubo,BL}. For the static XAS of Eu $M_5$-edge, the measurements were performed at 40 K by the total electron yield (TEY) mode with an ammeter.
Because the response time of an ammeter is larger than the order of ps, the partial electron yield (PEY) mode was instead used with a microchannel plate (MCP) detector for Tr-XAS \cite{Takubo}.

In the first step, the static XAS measurement of EuNi$_2$(Si$_{0.21}$Ge$_{0.79}$)$_2$ was performed using the TEY mode.
Figure 1 shows the Eu $M_5$-edge TEY XAS spectrum at 40 K.
It can be observed that the arrowed peaks are at 1129.6 eV and 1131.5 eV, which correspond to the main peaks of Eu$^{2+}$ and Eu$^{3+}$, respectively.
The shape of the spectrum is in good agreement with that in Ref. \cite{Yamamoto2}, indicating the fine quality of the fractured sample surface.

\begin{figure}
\begin{center}
\includegraphics[width=\linewidth]{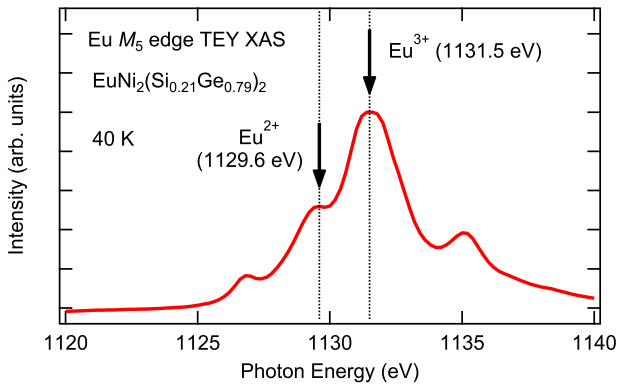}
\caption{(Color online) Eu $M_5$-edge TEY XAS spectra of EuNi$_2$(Si$_{0.21}$Ge$_{0.79}$)$_2$ at 40 K. Arrows indicate the characteristic peaks of Eu$^{2+}$ at 1129.6 eV and Eu$^{3+}$ at 1131.5 eV.}
\end{center}
\end{figure}

Based on the result of static XAS, Tr-XAS measurements were performed using the pump-probe technique.
The schematic diagram of the experimental setup is shown in Fig. 2(a).
A Ti:sapphire laser pulse ($h\nu=1.5$ eV, repetition rate $=1$ kHz, width $=50$ fs)  was adopted as the pump light for this study.
As a probe light, a synchrotron soft X-ray pulse (an isolated bunch in the
H-mode of SPring-8: a single bunch $+$ a bunch train)
with energy near Eu $M_5$-edge was used.
The spot size of the X-rays and Ti:sapphire laser is
$\sim$100 $\mu$m $\times$ 5 $\mu$m and $\sim$600 $\mu$m $\times$ 600 $\mu$m, respectively.
As the area irradiated by synchrotron X-ray is fully exposed by the laser irradiation, the dynamics induced by the pure laser incidence were probed.
By varying the delay time between the X-ray and the laser pulse, time-resolved information could be obtained \cite{Ogawa}.
A time resolution of 50 ps was decided by considering the width of the synchrotron X-ray pulse. 

For evaluating the time-resolved change of the XAS intensity, an oscilloscope was used.
The MCP output on the oscilloscope during the time-resolved measurements is shown in Fig. 2(b).
The solid and dotted lines correspond to the conditions of ``X-ray: ON, laser: ON'' and ``X-ray: OFF, laser: ON'', respectively.
The solid line indicates the bunches in H-mode detected mainly as photoelectrons from the sample.
On the other hand, the dotted line suggests the background originated from laser irradiation (such as stray light) and/or electronic noises.
The variations of voltage at $\sim$ $-3000$ ns and $\sim$ 100 ns were considered to be noises associated with the laser irradiation.
To remove this background noise, the dotted line was subtracted from the solid line.
After the removal of the background, the intensity of the XAS was obtained from the area of the pumped bunch (after laser irradiation) and the reference bunch (before laser irradiation).
Then, the rate of change was calculated by dividing the area of the pumped bunch by that of the reference bunch.
It should be noted that the sample quality (broken or not) can be verified by evaluating the reference bunch.

\begin{figure}
\begin{center}
\includegraphics[width=\linewidth]{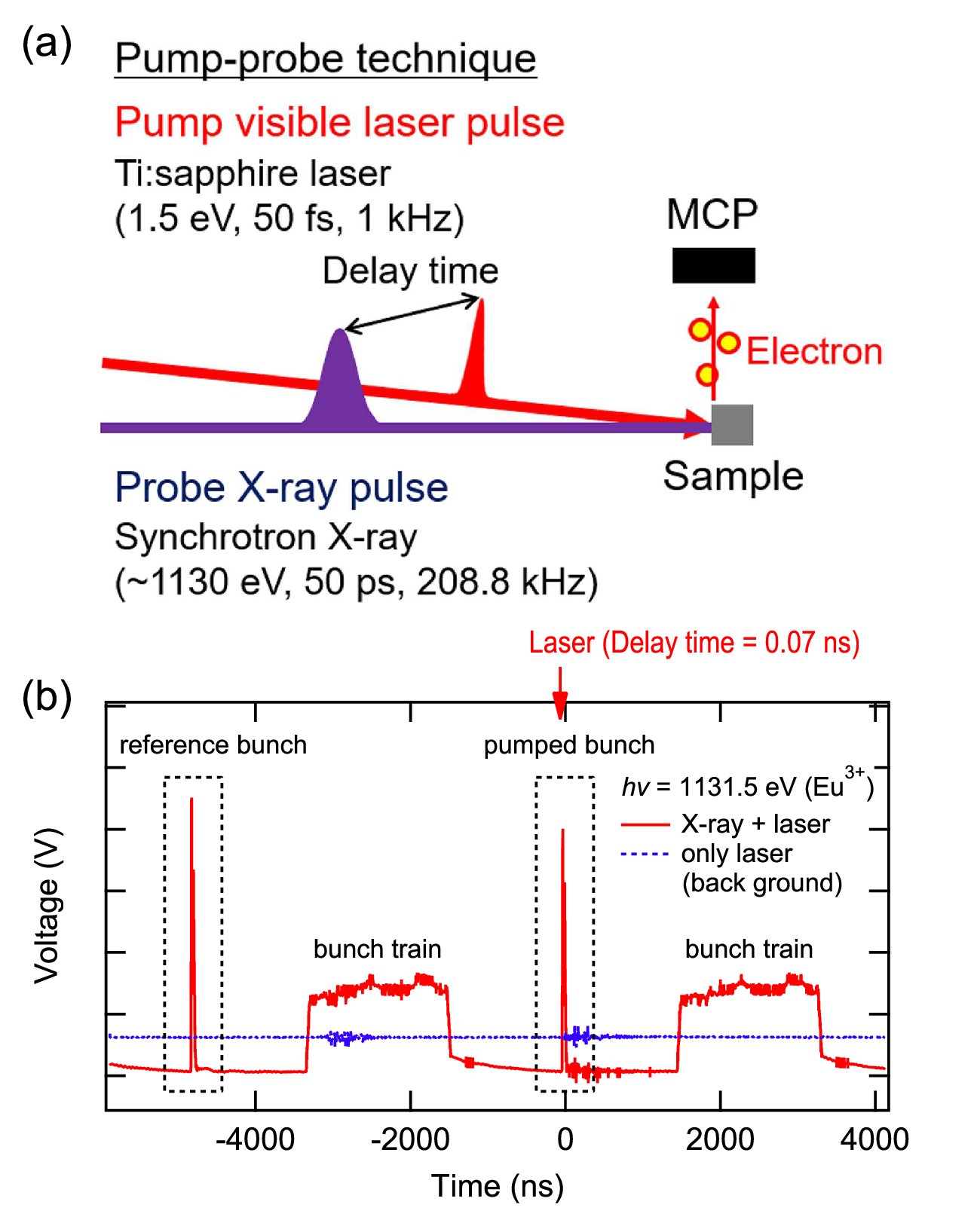}
\caption{(Color online) (a) Schematic diagram of setup for the pump-probe technique. The incident visible laser and X-ray pulse are overlapped at the sample surface. The delay time is defined as the time difference between the visible pump laser pulse and the probe X-ray pulse. (b) MCP output on the oscilloscope during the time-resolved measurements at a delay time $=$ 0.07 ns. The incident X-ray energy was set to 1131.5 eV (Eu$^{3+}$). Solid and dotted lines correspond to ``X-ray: ON, laser: ON'' and ``X-ray: OFF, laser: ON'', respectively.
The arrow indicates the laser position. The results are integrated over several thousand iterations.}
\end{center}
\end{figure}

\begin{figure}
\begin{center}
\includegraphics[width=\linewidth]{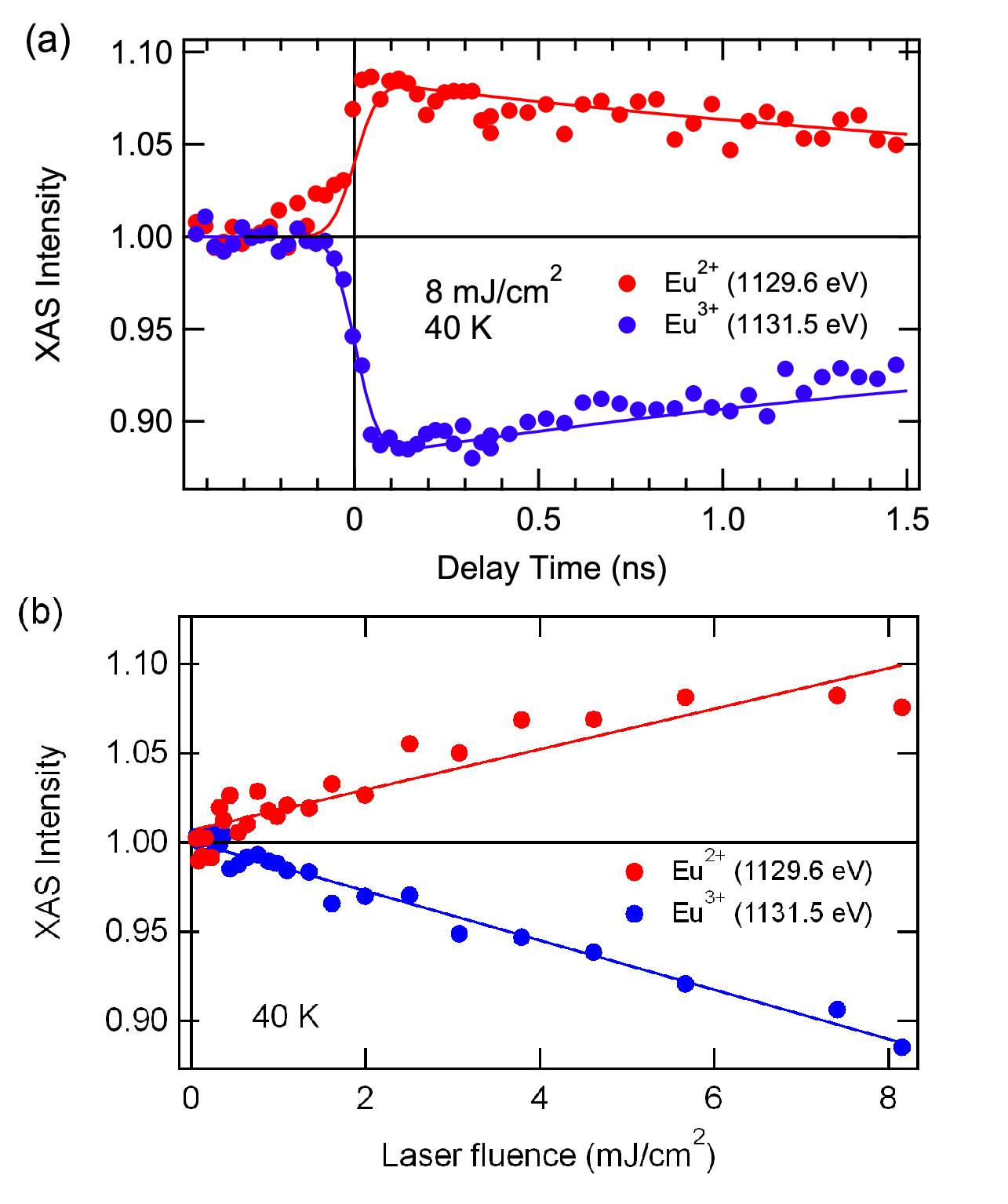}
\caption{(Color online) (a) Tr-XAS intensity for Eu$^{2+}$ (1129.6 eV) and Eu$^{3+}$ (1131.5 eV) obtained by the PEY mode. Dots and lines represent the experimental and fitting results. The measurements were performed at 40 K under the 1.5-eV laser irradiation with a fluence of $\sim$6mJ/cm$^2$. Delay time is defined as the time difference between the synchrotron X-ray bunch and the laser pulse. (b) Laser fluence dependence of XAS intensity in EuNi$_2$(Si$_{0.21}$Ge$_{0.79}$)$_2$ probed for Eu$^{2+}$ (1129.6 eV) and Eu$^{3+}$ (1131.5 eV) obtained by the PEY mode. During the measurements, the delay time was fixed at 0.07 ns. Dots represent the experimental results and lines denote the fitting results.}
\end{center}
\end{figure}

Figure 3(a) illustrates the Tr-XAS intensity probed at the energy for Eu$^{2+}$ (1129.6 eV) and Eu$^{3+}$ (1131.5 eV).
The sharp increase in intensity of XAS for Eu$^{2+}$ was observed by $\sim$10\% after a delay $=0$ ns, which suggests an increase in the ratio of Eu$^{2+}$ by the laser irradiation.
On the other hand, the intensity of Eu$^{3+}$ decreased by $\sim$10\% just after the laser irradiation.
These results suggest the mean valence of Eu becomes closer to Eu$^{2+}$ by laser irradiation, indicating a photoinduced valence transition between Eu 4$f$ and its conduction electrons.
The timescale of photoinduced valence dynamics was also evaluated using the following function
\begin{equation}
\begin{split}
I(t)&=I_1\exp(-t/\tau_{change})+I_2[1-\exp(-t/\tau_{recovery})]\\
&\quad+1-I_1.
\end{split}
\end{equation}
Because the time resolution of Tr-XAS is $\sim$50 ps,
the function was convoluted with the Gaussian response function ($\tau_{Gauss}=50$ ps).
From the fitting results represented by solid lines in Fig. 3(a), $\tau_{change}$ was found to be faster than 50 ps and $\tau_{recovery}$ was $\sim$3 ns.
As the timescale of valence transition was much longer than
a few ps, reported for YbAgCu$_4$ and YbInCu$_4$ via photoinduced reflectivity measurements \cite{Zhang},
a photo-induced metastable state may be realized in EuNi$_2$(Si$_{0.21}$Ge$_{0.79}$)$_2$.

\begin{figure}
\begin{center}
\includegraphics[width=\linewidth]{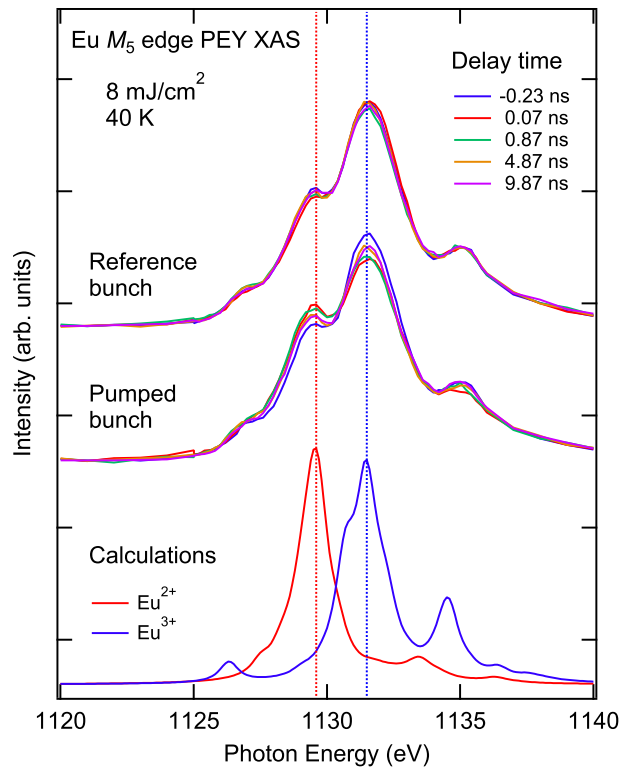}
\caption{(Color online) Eu $M_5$-edge PEY XAS spectra of EuNi$_2$(Si$_{0.21}$Ge$_{0.79}$)$_2$ at various delay times. The peaks at 1129.6 eV and 1131.5 eV correspond to the main peaks of Eu$^{2+}$ and Eu$^{3+}$, respectively.
The theoretical spectra were obtained by intra-atomic multiplet calculations of Eu$^{2+}$ and Eu$^{3+}$ from Ref. \cite{Yamamoto1}.}
\end{center}
\end{figure}

The laser fluence dependence was also investigated.
For those measurements, the delay time was fixed at 0.07 ns because the biggest changes were observed at that delay time.
Figure 3(b) shows the laser fluence dependence of XAS intensity probed at the peak energies of Eu$^{2+}$ (1129.6 eV) and Eu$^{3+}$ (1131.5 eV).
The fluence dependence can be fitted by a straight line, and there is no threshold,
such as the photoinduced insulator-to-metal transition \cite{Tsuyama},
which indicates that the valence of Eu can change by weaker laser fluence.
However, the valence change by temperature should have a threshold
because the valence of Eu changes drastically near the transition temperature, as shown in Ref. \cite{Ichiki}.
Therefore, a unique photoinduced valence transition was considered.

To investigate the photoinduced valence dynamics in further detail, the PEY XAS spectra were measured at different delay times.
Figure 4 illustrates the Eu $M_5$-edge PEY XAS spectra at delay times $=-$0.23 ns, 0.07 ns, 0.87 ns, 4.87 ns, and 9.87 ns.
The peaks at 1129.6 eV and 1131.5 eV correspond to the main peaks of Eu$^{2+}$ and Eu$^{3+}$, respectively. As for the reference bunches, the spectra were almost the same, indicating that the samples did not deteriorate during the measurements.
By changing the delay times between the laser and X-ray pulses, the shapes of XAS spectra can be changed, which corresponds to the photoinduced valence dynamics between Eu$^{2+}$ and Eu$^{3+}$.
Later, the mean valence of Eu was evaluated using the theoretical spectra obtained by the intra-atomic multiplet calculations
of Eu$^{2+}$ and Eu$^{3+}$ from Ref. \cite{Yamamoto1}.

Figure 5 (a) shows the comparison of experimental Eu $M_5$-edge PEY XAS spectra at various delay times with the theoretical spectra.
The valence of Eu was evaluated by comparing it with the linear combinations of the theoretical valences for Eu$^{2+}$ and Eu$^{3+}$.
As shown in the figure, the theoretically reproduced spectra were in good agreement with the experimental ones.
The evaluated valence of Eu at various delay times are shown in Fig. 5(b).
Before the laser irradiation, the valence of Eu was 2.76+.
The value is almost the same as that in Ref. \cite{Ichiki}.
Upon the irradiation of the laser pulse, the valence changes to 2.67+ at the delay time $=$ 0.07 ns, indicating photoinduced valence transition.
Then, the valence gradually recovered to its initial value, as shown in the figure.

\begin{figure}
\begin{center}
\includegraphics[width=\linewidth]{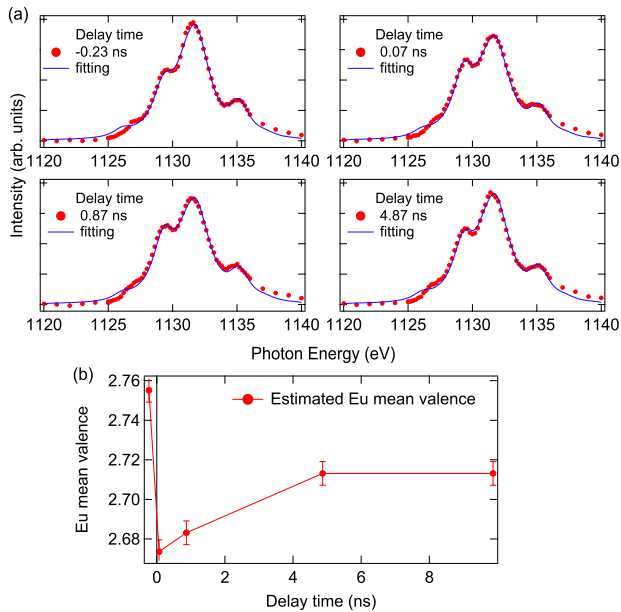}
\caption{(Color online) (a) Comparison between the experimental Eu $M_5$-edge PEY XAS spectra of EuNi$_2$(Si$_{0.21}$Ge$_{0.79}$)$_2$ at various delay times (dotted lines), and the theoretical spectra reproduced by the linear combinations of Eu$^{2+}$ and Eu$^{3+}$ (solid lines). (b) The time-dependent variation of mean valence of Eu evaluated by the calculations.
The error bars are obtained from the variation of the reference bunch spectra in Fig.4.}
\end{center}
\end{figure}

The observed Tr-XAS intensity in Fig. 3(a)
was interpreted as the variation of Eu valence via XAS spectrum at each interval of the delay time.
By comparing the XAS spectra at several delay times with the theoretical spectra, the photoinduced valence transition from 2.76$+$ to 2.68$+$ was revealed.
By considering a lifetime as long as $\sim$3 ns, and a valence change smaller than that induced by external stimuli such as temperature and magnetic field, the photoinduced valence transition was considered to be a different phenomenon from the conventional valence transition, and a photoinduced metastable state can hence be realized.

Static XAS and Tr-XAS measurements of EuNi$_2$(Si$_{0.21}$Ge$_{0.79}$)$_2$ were performed using native equipment at BL07LSU of SPring-8.
From the static XAS, the energies of the characteristic peaks of Eu$^{2+}$ at 1129.6 eV and Eu$^{3+}$ at 1131.5 eV were observed.
In the Tr-XAS measurements, the photoinduced valence dynamics between Eu$^{2+}$ and Eu$^{3+}$ were detected successfully.
The timescale of valence dynamics were evaluated, and the timescales for change and recovery were found to be shorter than 50 ps and longer than 3 ns.
The laser fluence dependence was also measured.
In the absorption spectra observed at different delay times, the mean valence of Eu was evaluated using the theoretical spectra, revealing a valence change of $\sim$0.1.
These characteristics are different from the valence transition by external stimuli such as temperature, indicating the novel photoinduced states.

\section*{Acknowledgement}
This work was carried out by the joint research in the Synchrotron Radiation Research Organization and the Institute for Solid State Physics, the University of Tokyo
(Proposal No. 2017A7403 and No. 2017B7403).

\end{document}